\documentclass[12pt,preprint]{aastex}

\shorttitle{Torus size of NGC~4151}
\shortauthors{Pott et al.}

\begin{document}

\title{Luminosity-variation independent location of the circum-nuclear, hot dust in NGC~4151}


\author{Jorg-Uwe Pott\altaffilmark{1,2,3}, Matt A. Malkan\altaffilmark{2}, Moshe Elitzur\altaffilmark{4}, Andrea M. Ghez\altaffilmark{2,5}, Tom M. Herbst\altaffilmark{1}, Rainer Sch{\"o}del\altaffilmark{6}, Julien Woillez\altaffilmark{3}}

\altaffiltext{1}{current address: Max-Planck-Institut f{\" u}r Astronomie, K{\" o}nigstuhl 17, D-69117 Heidelberg, Germany\newline {\tt jpott@mpia.de, herbst@mpia.de}}
\altaffiltext{2}{Div. of Astronomy \& Astrophysics, University
of California, Los Angeles,  CA 90095-1547\newline {\tt malkan@astro.ucla.edu, ghez@astro.ucla.edu}}
\altaffiltext{3}{W.~M.~Keck~Observatory, California Association for Research in Astronomy, Kamuela, HI 96743\newline {\tt jwoillez@keck.hawaii.edu}}
\altaffiltext{4}{Dept. of Physics and Astronomy, University of Kentucky, Lexington, KY 40506-0055\newline {\tt moshe@pa.uky.edu}}
\altaffiltext{5}{Institute of Geophysics and Planetary Physics,
	University of California, Los Angeles, CA 90095-1565}
\altaffiltext{6}{Instituto de Astrofisica de Andalucia - CSIC, Camino Bajo de Huetor, 50, 18008 Granada, Spain\newline {\tt rainer@iaa.es}}


%
\begin{abstract}
After recent sensitivity upgrades at the Keck Interferometer (KI), systematic interferometric 2~$\mu$m studies of the innermost dust in nearby Seyfert nuclei are within observational reach. Here, we present the analysis of new interferometric data of NGC~4151, discussed in context of the results from recent dust reverberation, spectro-photometric and interferometric campaigns. { The complete data set gives a complex picture, in particular the measured visibilities from now three different nights appear to be rather insensitive to the variation of the nuclear luminosity.} KI data alone indicate two scenarios: the $K$-band emission is either dominated to $\sim 90\%$ by size scales smaller than 30~mpc, which falls short of any dust reverberation measurement in NGC~4151 and of theoretical models of circum-nuclear dust distributions. Or contrary, and more likely, the $K$-band continuum emission is dominated by hot dust ($\gtrsim 1300~K$) at linear scales of about 50~mpc. The linear size estimate varies by a few tens of percent depending on the exact morphology observed. { Our interferometric, deprojected centro-nuclear dust radius estimate of $55\,\pm\,5$~mpc} is roughly consistent with the { earlier published} expectations from circum-nuclear, dusty radiative transfer models, and spectro-photometric modeling. { However, our data do not support the notion that the dust emission size scale follows the nuclear variability of NGC~4151 as a $R_{\rm dust}\,\propto\,L_{\rm nuc}^{0.5}$ scaling relation.} Instead variable nuclear activity, lagging,  and variable dust response to illumination changes need to be combined to explain the observations. 

\end{abstract}




\section{Introduction}
\label{sec:1}

Nearby Active Galactic Nuclei (AGN) are rosetta stones to understand the astrophysics close to an actively accreting super-massive black hole. 
Sub-parsec resolution is required to identify answers to the key questions of AGN physics: (i) How does galactic material flow down to the accretion disk to feed  luminosities close to the Eddington limit? (ii) Which role do play outflows and jets in the energetic interconnection between the AGN and its host? and (iii) Are AGN of different luminosity intrinsically similar? 
Studying the spectral energy distribution of quasars and type 1 AGN from the optical to the far-infrared reveals remarkably similar SED shapes, suggesting similar physics \citep{1986ApJ...308...59E,1989ApJ...347...29S,1993ApJ...404...94K,2006A&A...457...61R}. 
Therefore, resolving astrophysical phenomena in nearby AGN enables to better understand farther, unresolved nuclei, and to gauge correlations between nuclear luminosity and properties of the surrounding circum-nuclear environment. Being one of the brightest AGN on the sky, \object{NGC~4151} reveals this unique potential of nuclei closer than a few tens of Mpc's in numerous publications.

In this article, we concentrate on the origin of the near-infrared (NIR) emission of \object{NGC 4151}, and its relation to the nuclear luminosity.
NIR-AGN emission is characterized by a steep decline from optical
wavelengths down to about 1~$\mu$m, and on the long wavelength side by the quick rise of
a separate IR emission bump peaking at about 3~$\mu$m \citep{1986ApJ...308...59E,1993ApJ...404...94K}. The origin of this NIR excess
has been discussed throughout the past four decades \citep[probably one of the first were][]{1968AJ.....73..870P, 1969Natur.223..788R}. While it is not in question that the
original power originates in the accretion disk, it is uncertain if the NIR emission excess
derives directly from nuclear emission processes, or if mostly dust outside the broad line region (BLR)
re-radiates the nuclear emission \citep{1986ApJ...308...59E,1988Natur.336..749E,1987ApJ...320..537B,1992ApJ...400..502B}.
The proper measurement of the
nuclear near-infrared emission (and its discrimination from dust-re-processed light) is crucial
to derive the intrinsic SED of accreting supermassive black holes (SMBH) in the centers of
galaxies and to understand the radiative processes involved.

Even at the relatively detailed linear scale of  NGC~4151 ($\sim\,82\,{\rm mpc\,mas^{-1}}$, footnote~$^{(a)}$ in Table~\ref{tab:33}), the accretion disk itself and the surrounding BLR are too small to be spatially resolved by the current generation of optical or infrared telescopes or interferometers \citep{2006ApJ...644..133B}. 
However, the circum-nuclear dust distribution is now within reach of observations with infrared telescope arrays at sub-50~mas angular resolution \citep{2003ApJ...596L.163S,2004A&A...418L..39W,2004Natur.429...47J}.
Mid-infrared data of the VLT interferometer typically find in Seyfert nuclei 10~$\mu$m emission sizes of a few parsecs \citep{2009A&A...502...67T, 2009ApJ...705L..53B, 2009MNRAS.394.1325R}.
Dust plays an exceptional role among the circum-nuclear components.
It is not only assumed to significantly contribute to the near-to-mid-infrared continuum radiation of an AGN.
Also, a non-spherically symmetric dust distribution is widely assumed to explain the type~1~/~2 dichotomy of AGN emission line spectra. 
Radiative transfer models of clumpy distributions of dust clouds extending out of the equatorial plane are currently the favored explanation of the steadily increasing amount of observational data \citep{1987ApJ...320..537B,1992ApJ...400..502B,2005A&A...437..861S,2006A&A...452..459H,2008ApJ...685..147N,2008ApJ...685..160N,2009arXiv0909.4539H}. In the following we refer to the dust distribution as {\it torus}, keeping in mind that its detailed morphology is rather uncertain, might not resemble closely a smooth torus. 

It is currently an open debate, how the findings and models at MIR wavelengths connect to the innermost hot dust, which is expected to contribute to the continuum emission at 2~$\mu$m. \citet{2009ApJ...705..298M} argue for a spatially and chemically distinct inner component, based on $2\,-\,35\,\mu$m spectra. Also, the interferometric MIR data of NGC~4151 appear to reject that the nuclear flux at 10~$\mu$m is dominated by the outer, cooler part of the same structure, which dominates the 2~$\mu$m continuum \citep{2009ApJ...705L..53B}. Such results discourage simple attempts to unify the dust emission structures around AGN, and suggest multi-component models, or at least a significant change of dust composition and grain size distribution with the radial distance to the central engine \citep{2005A&A...437..861S, 2009A&A...493L..57K}.

We report in this article on first results of a new campaign using the Keck Interferometer (KI) to add observational constraints to the origin of the NIR continuum of NGC~4151.
The 85~m baseline and the sensitivity of the KI is adequate to resolve the linear distances of order 30-200 mpc around NGC~4151 which matches the theoretically calculated dust sublimation radii for this source. 
Recent sensitivity improvements of the KI \citep[][\footnote{Current KI performance numbers are given at the website of NASA Exoplanet Science Institute ({\sc Nexsci}): {\tt http://nexsci.caltech.edu/software/KISupport/v2/v2sensitivity.html}}]{2006SPIE.6268E..21W, 2008SPIE.7013E..10R} enabled us to repeat the early Swain et al. KI measurement, although the variable nucleus of NGC~4151 has been at a significantly fainter state during the time of observation.   
After a description of the observations (Section~\ref{sec:2}), we discuss our findings and the implications on the interpretation of the combined visibility data set { \citep[from][and the here presented observations]{2003ApJ...596L.163S, 2009A&A...507L..57K} } in Section~\ref{sec:5}. 
{ We concentrate our discussion on issues left open by the previous high resolution NIR observations of the AGN: (i) is the $K$-band emission rather dominated by unresolved accretion disk emission \citep[as favored by ][]{2003ApJ...596L.163S}, or by resolved circum-nuclear dust emission \citep{2009A&A...507L..57K}? (ii) How closely related are accretion disk luminosity and circum-nuclear dust location in NGC~4151 \citep{2009ApJ...700L.109K}? (iii) Is the currently described scatter between direct interferometric and indirect reverberation measurements of the torus size a systematic offset reflecting that the respective method probes slightly different dust reservoirs, or does the scatter rather reflect the currently achievable accuracy of either method in deriving the actual dust location \citep{2009ApJ...700L.109K,2009A&A...507L..57K}? }
The concluding remarks in Section~\ref{sec:6} summarize our current results and give a brief outlook on the project.

\section{Observations}
\label{sec:2}

\begin{table}
\begin{center}
\caption{\label{tab:4}
Observing log.}
\begin{tabular}{ccccccc}
\tableline\tableline
Target & date (UT) & H.A.$\,^{a}$ & $u,v$$\,^{b}$ & proj. B$\,^{b}$ & calibrators & $V^2$ \\
 && $[$hr$]$ &$[{\rm m}]$ & $[$m,\,deg(EofN)$]$  & [from Tab~\ref{tab:3}]& [calib.]$\,^{c}$ \\
\tableline
NGC~4151 & Dec.15,2008 & -1.3 & (56.4, 51.6) & (76.4, 47.5) & 1,2 & 0.85 \\
         & Dec.15,2008 & -1.1 & (56.0, 54.0) & (77.8, 46.0) & 1,2 & 0.84 \\
\tableline

\tableline
\tableline
\end{tabular}
\tablenotetext{a}{Hour angle}
\tablenotetext{b}{The {\it u,~v}-coordinates, given here in meters, are the baseline length (B) projected onto the line of sight. {\it u} points East, {\it v} points North. They are equivalent to polar values of the projected baseline, given in the next column in meters, and degrees East of North.}
\tablenotetext{c}{The {\it absolute} calibration accuracy is at least 0.03, which is shown in Fig.~\ref{fig:1} and should be used if comparing these values with results from different nights and observing campaigns. The {\it differential} intranight visibility precision is 0.01~-~0.015, based on the statistical scatter of the measurements.}
\end{center}
\end{table}

\begin{table}
\begin{center}
\caption{\label{tab:3}
Properties of the interferometric calibrator stars used for calibration of the instrumental transfer function during data reduction. }
\begin{tabular}{clccc}
\tableline\tableline
\# &Calibrator       & $V/H/K$$\,^{(a)}$ & Spec. Type$\,^{(a)}$ & Ang. diameter (mas)$\,^{(b)}$  \\
\tableline
1 & HIP58819         & 10.9/8.5/8.4      & K0III                & $0.11\,\pm\,0.03$ \\
2 & HD109691         & 8.9/8.9/8.9       & A0V                  & $0.04\,\pm\,0.02$ \\
\tableline \tableline
\end{tabular}
\tablenotetext{a}{from {\sc Simbad}}
\tablenotetext{b}{bolometric diameter fit from the {\sc Nexsci} getCal tool.}
\tablecomments{Calibrator stellar diameters significantly smaller than 0.5~mas, are unresolved by the KI. 
The statistical errors given here for the bolometric diameter fit to a black body likely slightly underestimate systematic errors of the NIR diameter of stars, but even 0.2~mas uncertainties in the diameter would not change the visibility calibration.}
\end{center}
\end{table}

The 85~m baseline of the Keck Interferometer (KI) is oriented 38$^\circ$ east of north (Colavita et al. 2004; Wizinowich et al.
2004). We used the $K'$-band (2-2.4~$\mu$m) $V^2$ continuum mode. All data shown here are from the white-light channel of the beam combiner, observed at an effective wavelength of 2.18~$\mu$m.  The observations were conducted on Dec.~15,~2008 (UT), details appear in Table~\ref{tab:4}. We followed standard observing and data reduction procedures \citep[see Section~3.2 in ][and references therein]{2003ApJ...592L..83C}. KI data are provided to the observer in a semi-raw state, and still require estimation and correction for the system visibility (or visibility transfer function), which is estimated by observing unresolved calibrators, close in space  and time (typical numbers for {\it close} are $\lesssim \, 10^\circ$ and $\lesssim \, 0.5\,{\rm hr}$).  
{ We chose calibrators of a $K$-band magnitude similar to the target, to avoid a flux bias in the calibrated visibilities ($V^2$, Table~\ref{tab:3})}. 
Our typical scanlength, leading to one visibility point, is 200~sec.

A reliable data calibration is indicated by stable system visibility and the flux ratios between both telescope beams over the night. Our observation was conducted at a visible seeing of 0.5".
To estimate and apply the system visibility, we used the wb/nbCalib-software suite by {\sc Nexsci}.
We use the standard deviation over the about 25 individual visibility estimations, which make up the 200~s scan, as a first estimate of the uncertainties of each data point. The resulting statistical noise in the calibrated visibility measurement is 0.01-0.015, proving the good observing conditions of the night. This gives the differential intranight visibility precision, and can be used to check for visibility slopes with respect to changes of the baseline projection.
Long-term monitoring of KI data against known binary orbits shows that the absolute calibration accuracy of the data from night to night is at least 0.03 \footnote{{\tt http://nexsci.caltech.edu/software/KISupport/v2/v2sensitivity.html}}. The difference between the intranight precision, and internight accuracy is visualized in Fig.~\ref{fig:1}. We use the 0.03 uncertainty to derive absolute emission sizes (Table~\ref{tab:33}), and to check for flux-induced size variations (Fig.~\ref{fig:11}).

{ The NIR flux of NGC~4151 is variable, and a contemporaneous total flux measurement is needed in our discussion of the data, to properly compare our visibilities with data from other observing epochs.
A direct flux calibration of the total photon counts detected by the interferometric camera (called {\sc Fatcat}) needs to reflect, that {\sc Fatcat} is fed by single-mode fibres. The fibres realize a spatial filtering of $\sim\,50~{\rm mas}$ on sky, to reduce the atmosperic phase noise. Since this quasi-aperture is on the order of the Keck diffraction limit at 2.2~$\mu$m, this introduces a Strehl dependence of the $K$-band flux measurements. The data show that the Strehl on NGC~4151 is systematically lower than on the calibrator stars, because the stars are significantly bluer then the red AGN and deliver more photons to the visible wavefrontsensor (WFS; note that we matched calibrator stars and AGN in the infrared brightness to get good visibility accuracy). Furthermore, the WFS probably detects some extended flux from the host galaxy. 

To account for this differential Strehl, the raw data of the AGN are multiplied by 1.45, before the flux calibration of the raw data against the calibrator stars.
 This empirical correction factor for the differential Strehl (between stars and AGN) of 1.45 appears to be stable over time, within the precision of the flux count estimation. It fits to the data of the two campaigns targeting NGC~4151, where quasi-simultaneous $K$-band photometry from a separate telescope is available: \citet[][]{2009A&A...507L..57K} and \citet{2003ApJ...596L.163S}. While the former publish quasi-contemporaneous UKIRT photometry, we reanalysed the archived KI-data of the latter paper, and compared the {\sc Fatcat} photometry against the simultaneous {\sc Magnum} single telescope photometry, as presented by \citet{2009ApJ...700L.109K}. To cross-check that by applying the correction factor, we account properly for the loss of light due to the spatial filtering of our $K$-band data, we also flux-calibrated the spatially unfiltered 1.6~$\mu$m ($H$)-band flux of the tip-tilt imager next to {\sc Fatcat}, which feeds a tip-tilt stabilizing loop during KI measurements. Both this $H$-band flux (50~mJy), and the corrected $K$-band flux (95~mJy) fit the corresponding values in \citet[][]{2009A&A...507L..57K}, with the {\it same} flux offset of a factor of 1.9, which we account to intrinsic flux variability.
 In Fig.~\ref{fig:11}, we used for the $K$-band photometry of the three visibility measurements the corrected {\sc Fatcat} photometry for the Swain et al., and our data, and the UKIRT photometry from Kishimoto et al.~2009, together with the respective photometric uncertainties. The flux contamination of the host galaxy in the $K$-band is with about 1~\% neglectable \citep{2009A&A...507L..57K}. The flux correction factor of 1.45 for the {\sc Fatcat} $K$-band data of NGC~4151 seems to be stable over the years, and therefore not too sensitive to the actual observing conditions, as long as a similar calibration scheme is used. 
 This is probably due to the fact, that in case of poor, unstable AO performance the interferometric measurement would not work efficiently.
 But the correction factor should not be applied to future datasets without careful checking of the effective differential Strehl between calibrator and target during the respective observation. In particular, the correction factor should be re-estimated for KI observations of different targets.
}

\section{Discussion of the results}
\label{sec:5}
\begin{figure}
\epsscale{.80}
\plotone{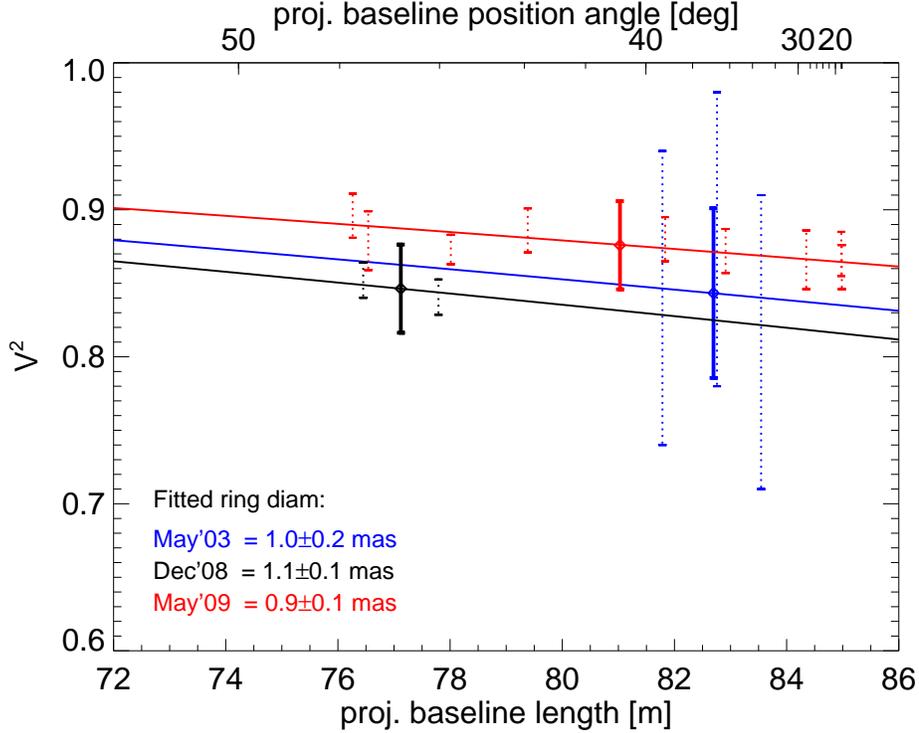}
\caption{\label{fig:1}
Calibrated visibility data from our observations (black), and from two other measurements \citep[blue, red:][respectively]{2003ApJ...596L.163S,2009A&A...507L..57K}, plotted versus the projected baseline length. The dotted errorbars show the individual data points, with their estimated  uncertainties. The three larger. solid errorbars show the mean $V^2$-result of each campaign, and the adequate accuracy bias of at least 0.03. The smaller statistical errors reflect the differential precision of the observations within a night, the larger accuracy bias needs to be taken into account if absolute size scales are fitted, and if observations from different nights are compared. $\chi ^2$ minimized face-on ring models are overplotted, the best fit diameters are given. A compact flux contribution from the AGN on the order of up to 25~\% does not alter the results significantly (see Table~\ref{tab:33} for various model fit results). { The total $K$-band flux shown comprises the flux contributions from the accretion disk and from the circum-nuclear dust (see Sect.~\ref{sec:2} for flux calibration details).}}
\end{figure}

\begin{figure}
\epsscale{.80}
\plotone{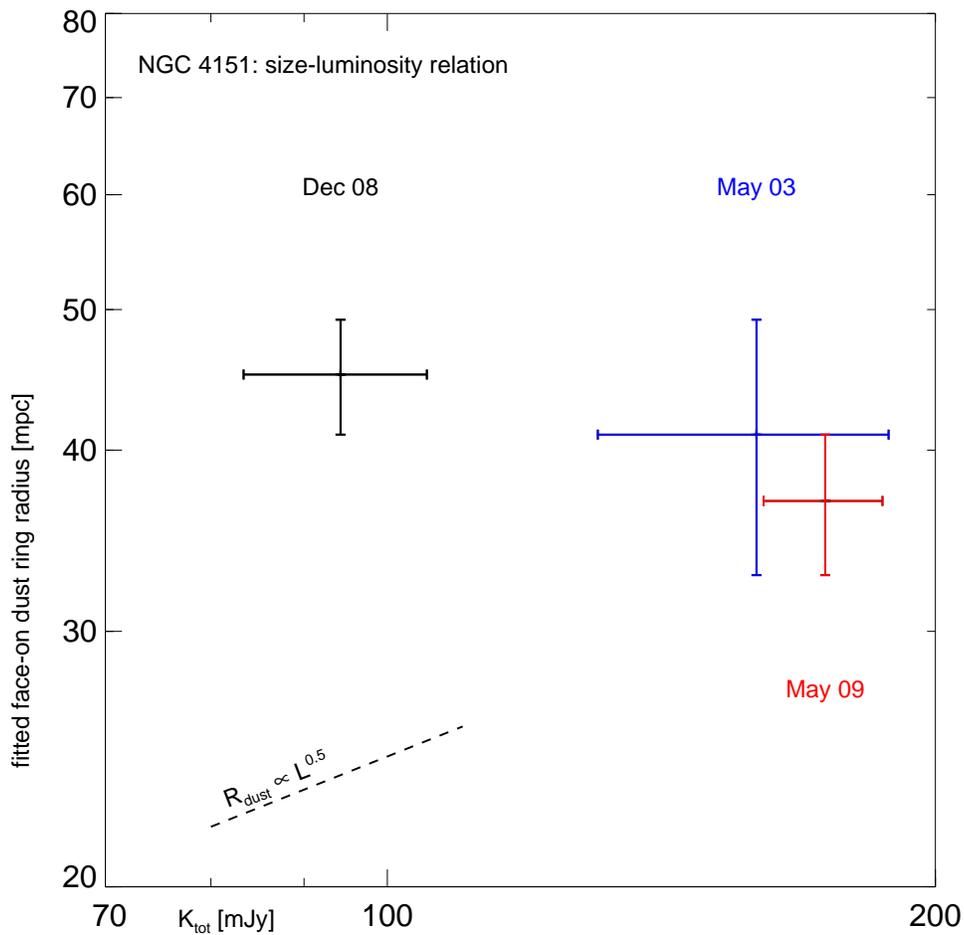}
\caption{\label{fig:11}
Dust ring radii, derived from fitting a face-on ring to the calibrated visibility data. Color-coding matches Fig.~\ref{fig:1}. In the lower left, a $L^{0.5}$-dependence is shown for comparison. Note, that the plotted radii neither take into account projection effects nor variable nuclear flux contributions. An inclined dust ring would have larger centronuclear radii. A nuclear flux contribution, varying between 10~-~25\%, could explain the apparent slope of the three datapoints, as discussed in Section~\ref{sec:33}.   }
\end{figure}

\begin{figure}
\epsscale{.98}
\begin{center}
\plotone{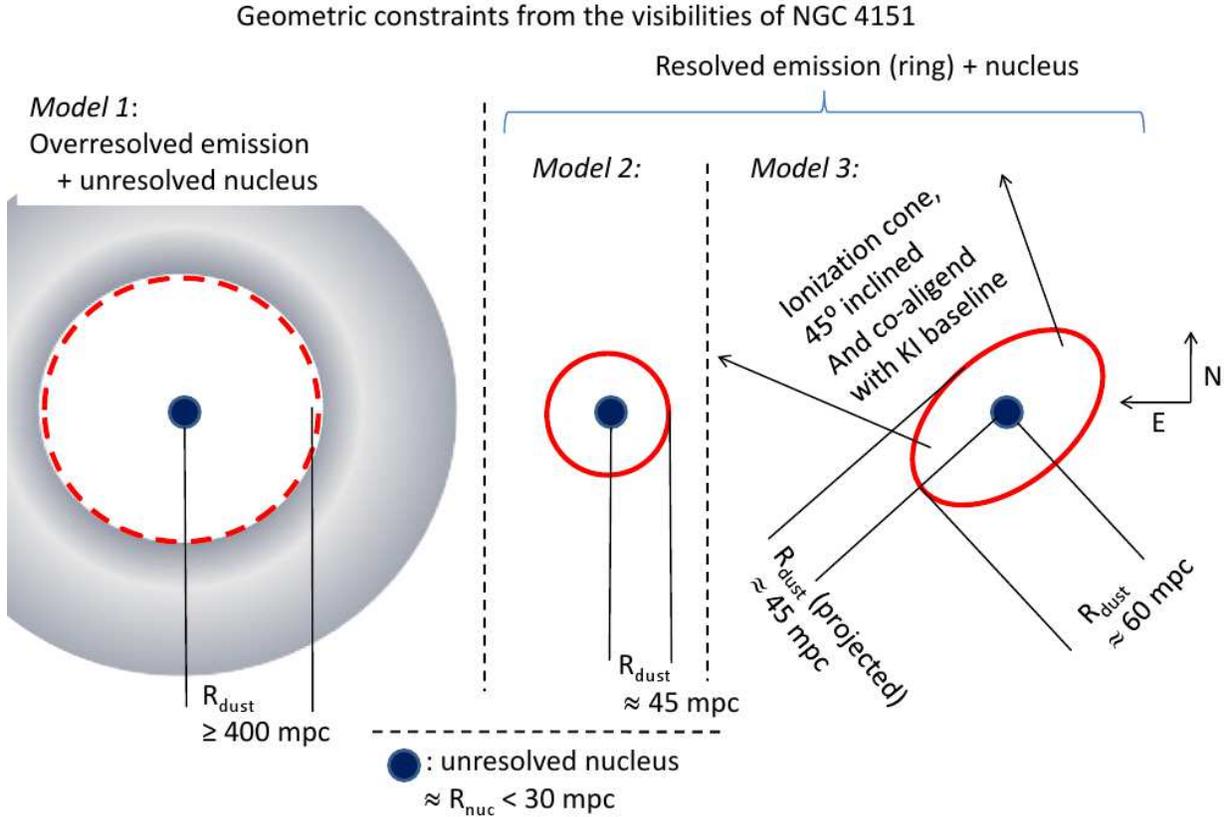}
\end{center}
\caption{\label{fig:2}
Sketches (not to scale) of the geometric constraints on the extended $K$-band emission of NGC~4151, seen through the 50~mas spatial filter of the Keck interferometer. Details are given in Section~\ref{sec:111}. Note, that the overresolved emission in model 1 might be diffuse, or more spherically than drawn, but the emission needs to origin from centronuclear radii larger than 400~mpc to be completely resolved out, and to not contribute to the $V^2$-budget.
}
\end{figure}

\begin{table}
\begin{center}
\caption{\label{tab:33}
Model fit parameters consistent with the data. The varying fit results of the ring models \#2 and \#3 reflect projection effects and represent general lower and upper limits on the intrinsic $R_{\rm dust}$ (Sect.~\ref{sec:111}).}
\begin{tabular}{ccccc}
\tableline\tableline
Model & $FR_{\rm nuc}^K$ & \multicolumn{2}{c}{$R_{\rm dust}$}      & Type of dust distribution  \\
& & [mas] & [mpc]~$^{(a)}$ & \\
\tableline
\#1 & 0.9  & $\gtrsim\,5 $&$\gtrsim\,400 $&overresolved \\
\#2 & 0.25 & $0.55\,\pm\,0.05$& 45 & resolved ring\\
\#2 & 0    & $0.6 \,\pm\,0.05$& 50 & resolved ring\\
\#2 & 0.25 & $0.45\,\pm\,0.05$& 35 & resolved Gaussian \\
\#2 & 0    & $0.5 \,\pm\,0.05$& 40 & resolved Gaussian \\
\#3 & 0    & $0.75\,\pm\,0.05$& 60 & resolved ring, 45$^\circ$ inclined,\\
    &      &                 &    & minor axis co-aligned with the KI baseline\\
\tableline \tableline
\end{tabular}
\tablenotetext{a}{from {\sc Ned}: Cosmology-corrected redshift of 0.00414 against the 3K~microwave~background reference frame results in angular-size distance of 16.9~Mpc, and 82~mpc/mas, respectively, with $H_0\,=\,73\,{\rm km\,sec^{-1}\,Mpc^{-1}}, \Omega_{\rm matter}\,=\,0.27$}
\end{center}
\end{table}

The calibrated visibilities are plotted in Fig.~\ref{fig:1} together with previously published data from \citet{2003ApJ...596L.163S, 2009A&A...507L..57K}. 
It is apparent that the different interferometry campaigns measured very similar visibilities, despite of the significantly different total flux measured (Fig.~\ref{fig:11}).
By fitting models, we focus on three different properties of the size constraints which can be derived from the visibilities: the amount of unresolved flux, the size of the resolved flux emission structure, and the deprojection of this size. Our models consist of a variable fractional amount of (unresolved) nuclear emission ($FR_{\rm nuc}^K\,=\,0\,-\,1$) along with resolved (overresolved) emission at (beyond) centro-nuclear radii $R_{\rm dust}$ (Fig.~\ref{fig:2}). We simply refer to the extended component as dust, being the most likely origin of it. \newline

\subsection{\label{sec:111}Modeling the dust emission}
{\it Model~1:} Already a single visibility below unity shows that some fraction of the flux is resolved by the interferometer. This corresponds to $FR_{\rm nuc}^K\,<\,1$. The most compact brightness distribution, still consistent with the data, would put $FR_{\rm nuc}^K\,\sim\,90~\%$ of the flux at radii smaller than 0.35~mas ($<$~30~mpc)  below the resolution limit of the KI ("point source"), and the remainder of the flux outside of $R_{\rm dust}\,\sim\,400~{\rm mpc}$ to be completely resolved out. 
Since radii of about 30~mpc are significantly smaller than the dust sublimation radius in NGC~4151{ , even for large graphite grains \citep[see the recent discussions in ][]{2007A&A...476..713K, 2008ApJ...685..160N}}, this scenario would suggest that the $K$-band emission is dominated by hot, and potentially partially non-thermal nuclear emission from the innermost accretion disk zone. 

But the derivation of $FR_{\rm nuc}^K$ from a few visibilities is degenerate. Reducing $FR_{\rm nuc}^K$ requires only to shrink $R_{\rm dust}$ to stay consistent with the data. The visibility alone does not constrain $FR_{\rm nuc}^K$ any further than being $\leq$~0.9. However, recent spectro-photometric and NIR-polarimetric data appear to limit $FR_{\rm nuc}^K$ to less than 25~\%. { In these studies,} the central $K$-band flux appears dominated by moderately warm thermal emission ($\sim\,1500~K$), likely from circum-nuclear dust clouds \citep[see the detailed discussions and references of the { observational support of dust being} the dominating origin of the $K$-band emission in Seyfert~1 nuclei in][]{1987ApJ...320..537B,2009ApJ...698.1767R,2009A&A...493L..57K,2009A&A...507L..57K}. 
{ Since our KI photometry and visibilities from the Dec~2008 run are consistent with these earlier }findings, we assume in the following $FR_{\rm nuc}^K\,\lesssim\,0.25$, which means that both total and correlated flux are dominated by extended emission with an emission size similar to the angular resolution of the interferometer.
Such resolved emission would result in decreasing visibilities with increasing baselinelength.
Indeed such a trend is marginally apparent in both new datasets with high intranight differential visibility precision { (Dec~08, and May~09 in Fig.~\ref{fig:1})}. 

{\it Model~2:} aims at estimating the order of magnitude of $R_{\rm dust}$. To translate visibilities into size scales, a simple model of the extended emission is required. Gaussians and rings fit the data equally well, and give the similar $R_{\rm dust, circ}$ of $\sim\,45$~mpc ($0.5\,\cdot\,{\rm FWHM}\,\approx\,R_{ring}\,=\,0.55\pm 0.1\,{\rm mas}$, Fig.~\ref{fig:1} and Table~\ref{tab:33}). This size estimate is largely unaffected by an $FR_{\rm nuc}^K\,\lesssim 25\%$, or by the radial extension of the ring as long as the ring radius is larger than its extension. The key difference between Gaussian and ring models is, that the Gaussian not only defines a size scale, but at the same time it constrains the emitting {\it area}.
The fitted FWHM translates into a solid angle of 0.92~mas$^2$. However, a brightness temperature calculation rejects a smooth Gaussian-like emission surface profile.
Assuming a black body temperature of 1300~K, this surface would result in about 500~mJy.
This is about a factor of 2.5-10 larger than the measured dusty $K$-band emission of NGC~4151, depending on its status of nuclear activity (Fig.~\ref{fig:11}). 
If the dust emission is optically thick, this discrepancy means that the radiating surface area is significantly smaller than the solid angle of the fitted Gaussian. If we were to shrink the FWHM of the Gaussian to match the observed fluxes, the resulting increased visibilities would overshoot over the data. Higher dust temperatures would strengthen this argument against a smooth Gaussian brightness distribution.
Thus we conclude that the black body component, apparent in the spectro-photometry \citep[e.g.][]{2009ApJ...698.1767R}, is not produced by an approximately spherical, smooth dust distribution (which would resemble a Gaussian in projection). The planar geometry of the ring model seems more reasonable. This matches the expectation for observing a type~1 AGN. Optically thick clouds, as used in modern torus models, would reduce the radiating surface as well (without changing the $R_{\rm dust}$ estimate). The ring morphology also appears reasonable due to the lack of significant extinction towards the very nucleus \citep{1982ApJ...256...75L}.

{\it Model~3:} The $R_{\rm dust, circ}$ estimate from the previous paragraph assumes a face-on, circular dust distribution. However, given the single KI baseline, the size constraints of KI-data are primarily one-dimensional along the position angle of the baseline (about 40$^\circ$). The resolved source could be substantially larger along the orthogonal (SE-NW) direction. In fact, the KI baseline position angle roughly co-aligns with the ionization cone axis of symmetry of NGC~4151.
\citet{2005AJ....130..945D,2009ApJ...698.1767R} report position, opening and line-of-sight inclination angles of the cone to be $\sim\,60^\circ$, $\sim\,70^\circ$, and $\sim\,45^\circ$. If the circum-nuclear dust, seen as extended component by the interferometer, is responsible for the formation of the ionization cone, the opening angle would translate into a torus half-height of 0.7 times the inner radius.
Such a geometry matches the thin ring model, if the extended $K$-band emission is dominated by the hottest dust exposed to direct nuclear illumination. { 
If KI and the projected minor axis were exactly co-aligned, the size estimate from the previous face-on model ($R_{\rm dust, circ}\,\sim\,45\,{\rm mpc}$) needs to be corrected for the inclination angle to $R_{\rm dust, incl}\sim\,60\,{\rm mpc}$ (see Table~\ref{tab:33}).
The position angle of the inner radio-jet, a possibly better constraint on the AGN axis of symmetry, lies with a PA of $\sim\,80^\circ$ within this ionization cone \citep{2003ApJ...583..192M}, but slightly off the KI position angle. Summarizing, the intrinsic 3d inner torus radius $R_{\rm dust}$ is constrained to $R_{\rm dust, circ}\,\le\,R_{\rm dust}\,\le\,R_{\rm dust, incl}$. Taking into account the orientation constraints of radio jet and ionization cone, we estimate $R_{\rm dust}\,=\,55\,\pm\,5\,{\rm mpc}$.
} The significant inclination of the nuclear axis of symmetry, indicated by the inclination of the ionization cone, is supported by the time-variable, spectral classification of NGC~4151 as Seyfert 1.5~-~1.8 \citep{2008A&A...486...99S}.

\subsection{\label{sec:33}
On the luminosity dependence of the torus size}

In general, both $V$-to-$K$-continuum reverberation and NIR interferometric measurements of nearby AGN appear to confirm a $L^{0.5}$-dependence of the extended $K$-band emission size on the nuclear illumination luminosity \citep{2006ApJ...639...46S,2009A&A...507L..57K}. In addition, the NIR SED bump of AGN can be fitted consistently with a black body profile of about 1500~K, for AGN samples covering orders of magnitude in nuclear power \citep{1993ApJ...404...94K}. These are strong arguments for the $K$-band being dominated by thermal dust emission close to the sublimation limit. The strong nuclear emission variability of NGC~4151 by several 100~\% on yearly timescales makes it an ideal target to do the next step, and to study the relation between the nuclear emission and the dust emission size in detail. 
Sequential dust reverberation studies in several AGN did find changes of the response delay time, suggesting a change of $R_{\rm dust}$ in reaction to different nuclear illumination \citep{1992ApJ...400..502B,2009ApJ...700L.109K}. 
However, given a $R_{\rm dust}$ of several tens of mpc, it would require unreasonable fast dust bulk motion (at 0.1~c) to follow the luminosity changes according to a $L^{0.5}$-law. 
Therefore, cyclic dust grain destruction and formation was suggested to explain both the observed infrared flux variations \citep{1992ApJ...400..502B} and the large scatter and relatively poor $L^{0.5}$-dependence in the dust reverberation size estimates of NGC~4151 \citep{2009ApJ...700L.109K}.

For the first time, we can compare these claims of a changing $R_{\rm dust}$ against a multi-epoch visibility dataset, taken at different brightness states (Fig.~\ref{fig:11}) of NGC~4151.
{ 
Assuming a simple $L^{0.5}$ scaling, the flux variation between the two new datasets ($F_{\rm K, \,May09}\,/\,F_{\rm K,\,Dec08}\,=\,1.9\,\pm\,0.3$) would result in a size increase by a factor of $1.4\,\pm\,0.1$.  Such large size variations do not fit the multi-epoch dataset, since the typical precision of the centronuclear radius estimate is $\lesssim\,10~\%$.
In addition, the flux-dependence of $R_{\rm dust}$ is constrained to $R\,\propto\,L^{(-0.3\,\pm\,-0.3)}$ by the data. Thus, if indeed we would observe a variation of $R_{\rm dust}$, then the trend would predict larger sizes at lower luminosity states, in contrast to a $L^{0.5}$-dependence.}

{ However, the apparent slope can be explained by $FR_{\rm nuc}^K$ variability alone, without the need for a variation of $R_{\rm dust}$ between the three observations.} 
\citet{2007A&A...476..713K} measured a significantly higher nuclear flux contribution to the $K$-band at a brighter state of the nucleus than during the observation of \citet{2009ApJ...698.1767R} (20-25\% with respect to $~$10\%), although both teams fitted in comparable ways very similar black body temperatures to the NIR-emission. 
The comparison of the two new precise visibility measurements reveal a small $V^2$-offset of 0.04 at the same angular resolution (compare the black and red data in Fig.~\ref{fig:1}. 
Such a difference would result from changing $FR_{\rm nuc}^K$ between both measurements from a 10~\% level in Dec~08 to the 20~\% level in May~09. The small $V^2-$difference, slightly larger than the conservative internight accuracy estimate of 0.03, can be accommodated by a change of the compact $FR_{\rm nuc}^K$, without further assumptions on the structure of the extended emission, in particular without changing the morphology of the extended brightness distribution.
However, a change of the thermal flux by a factor of two, without changing the size of the radiating surface, would require a temperature change by $\sim\,1.2\,.$ As far as we know, such a luminosity dependent dust temperature variation has not been observed yet in NGC~4151. In contrast, an increased radiation surface of a cloudy torus of fixed size would point to larger cloud covering factors or an increased number of efficiently heated clouds in brighter states. 

Optical polarization, and NIR-continuum variations appear correlated with visible nuclear continuum variations and enable reverberation experiments \citep[see][for a polarization reverberation experiment]{2007arXiv0711.1019G}.
Therefore, a variation of the fractional $FR_{\rm nuc}^K$ cannot be explained by a varying line-of-sight extinction towards the nucleus (e.g. due to passing clouds), because such an apparent nuclear luminosity change should not result in the observed correlated change of optical polarization, and NIR-continuum fluxes.
However, in a phase of nuclear brightening, the torus response might lag the nuclear flux rise by the reverberation delay. This could explain a momentary increase in $FR_{\rm nuc}^K$ and in $V^2$, even if the intrinsic $R_{\rm dust}$ does not change. 

$50\,{\rm mpc}$ equals 60 days of light travel, which matches the first four dust reverberation measurements of \citet{2009ApJ...700L.109K}. Because these earlier dust reverberation measurements span already a factor of six in optical luminosity, they also suggest a luminosity-variation independent dust emission size. The first KI measurement by \citet{2003ApJ...596L.163S} is quasi-contemporaneous with one of these dust reverberation measurements. Thus both methods appear to agree very well. However, the former authors report on four subsequent, significantly smaller reverberation size estimates, the shortest being $\sim\,$30~mpc only. 
Reverberation size estimates are biased to the innermost responding material \citep{1986ApJ...305..175G}. 
Furthermore, if the responding structure is rather flat and inclined with respect to the line-of-sight, as expected for NGC~4151, the first response can set in considerable faster than light travel time corresponding to $R_{\rm dust}$. However, in case of an isotropic illumination, and a fixed dust geometry, such geometric effects could only explain a fix offset between reverberation and interferometric radii.

This discussion implies that dust reverberation sizes may often underestimate the location of the bulk of the circum-nuclear dust, which dominates the $K$-band emission. Indeed, if we use our $R_{\rm dust}$ for NGC~4151, at least three of the four interferometric $R_{\rm dust}$ estimates of the AGN, observed by \citet{2009A&A...507L..57K}, exceed the $L^{0.5}$-fit of the reverberation size by factors of about $2\,\pm\,0.5$. The fourth object, Mrk~231, likely has an inclined geometry, and larger, de-projected $R_{\rm dust}$ as well \citep{2004MNRAS.350..140S}.
The direct interferometric $R_{\rm dust}$ estimate appears as the more robust measurement of the average location of the bulk of the warm circum-nuclear dust (if projection ambiguities are well constrained). { In contrast, the smallest reverberation sizes are smaller than the interferometric sizes, and the reverberation sizes vary beyond the scatter of the interferometric size estimates.
We suggest, that the reverberation sizes are probably tracing the location of momentary illuminated innermost, very hot dust clouds, instead of the bulk of the $K$-band dust.}
If circum-nuclear dust originates in outflows from the central engine \citep[][and references therein]{2006ApJ...648L.101E}, then the shortest dust reverberation times could derive from optically thick dust in such outflows inside $R_{\rm dust}$.

\section{Conclusions} 
\label{sec:6}

The results of a new interferometric NIR observing campaign of NGC~4151 are presented. Although this time in a 2 times fainter state, we managed to re-observe the AGN with the Keck Interferometer thanks to recent sensitivity improvements of the instrument.
The interferometric visibilities and size estimates were compared to previously published interferometric and single telescope data. 
The current KI datasets suggest that the major part of the radiation originates in a (possibly inclined) toroidal structure of an intrinsic, deprojected radius of about $55\,\pm\,5$~mpc, comparable to $V$-to-$K$-continuum reverberation measurements, without constraining the morphology in further detail.

Our dataset { and its comparison to published data enables for the first time to study the sensitivity and response of near-infrared interferometric visibilities of an AGN to intrinsic flux variations}. The observations show that NIR-interferometry on bright AGN ($K\,\sim\,$10~mag) are now feasible at a high precision of a few percent. We did not detect a significant visibility dependence on doubling the $K$-band emission of NGC~4151. This supports the notion, that, for an individual AGN, the size of its circum-nuclear NIR emission structure does not strictly depend on the respective momentary nuclear luminosity.  The direct interferometric size estimates appear to be more robust estimates of the average location of the dust than NIR continuum reverberation estimates. Being flux-independent, this average location of the hot dust probably relates to the sublimation radii of the AGN at its high activity state.
Future dust reverberation campaigns,  contemporaneous to direct interferometric measurements are highly desirable to study the circum-nuclear NIR emission region, to correctly interpret the reverberation measurements, and to investigate apparent changes in the dust illumination, heating efficiency, or covering factors.

\acknowledgments

We are grateful to the excellent KI team at WMKO and NExScI for making these observations a success. 
K. Meisenheimer and L. Burtscher contributed helpful discussions.
The data presented herein were obtained at the W.M. Keck Observatory, which is operated as a scientific partnership among the California Institute of Technology, the University of California and the National Aeronautics and Space Administration. The Observatory was made possible by the generous financial support of the W.M. Keck Foundation. The authors wish to recognize and acknowledge the very significant cultural role and reverence that the summit of Mauna Kea has always had within the indigenous Hawaiian community.  We are most fortunate to have the opportunity to conduct observations from this mountain.
The Keck Interferometer is funded by the National Aeronautics and Space Administration as part of its Navigator program.
This work has made use of services produced by the NASA Exoplanet Science Institute at the California Institute of Technology.
This research has made use of the SIMBAD database, operated at CDS, Strasbourg, France.
This research has made use of the NASA/IPAC Extragalactic Database (NED) which is operated by the Jet Propulsion Laboratory, California Institute of Technology, under contract with the National Aeronautics and Space Administration.



{\it Facilities:} \facility{Keck:I ()}, \facility{Keck:II ()}




\clearpage

\begin{thebibliography}{}
\bibitem[Barvainis(1987)]{1987ApJ...320..537B} Barvainis, R.\ 1987, \apj, 
320, 537 
\bibitem[Barvainis(1992)]{1992ApJ...400..502B} Barvainis, R.\ 1992, \apj, 
400, 502 
\bibitem[Bentz et al.(2006)]{2006ApJ...644..133B} Bentz, M.~C., Peterson, 
B.~M., Pogge, R.~W., Vestergaard, M., \& Onken, C.~A.\ 2006, \apj, 644, 133 
\bibitem[Burtscher et al.(2009)]{2009ApJ...705L..53B} Burtscher, L., Jaffe, 
W., Raban, D., Meisenheimer, K., Tristram, K.~R.~W., R{\" o}ttgering, H.\ 2009, \apjl, 705, L53 

\bibitem[Colavita et al.(2003)]{2003ApJ...592L..83C} Colavita, M., et al.\ 
2003, \apjl, 592, L83

\bibitem[Das et al.(2005)]{2005AJ....130..945D} Das, V., et al.\ 2005, \aj, 
130, 945

\bibitem[Edelson 
\& Malkan(1986)]{1986ApJ...308...59E} Edelson, R.~A., \& Malkan, M.~A.\ 1986, \apj, 308, 59
\bibitem[Edelson et al.(1988)]{1988Natur.336..749E} Edelson, R.~A., Gear, 
W.~K.~P., Malkan, M.~A., \& Robson, E.~I.\ 1988, \nat, 336, 749 
\bibitem[Elitzur 
\& Shlosman(2006)]{2006ApJ...648L.101E} Elitzur, M., \& Shlosman, I.\ 2006, \apjl, 648, L101 

\bibitem[Gaskell 
\& Sparke(1986)]{1986ApJ...305..175G} Gaskell, C.~M., \& Sparke, L.~S.\ 1986, \apj, 305, 175
\bibitem[Gaskell et al.(2007)]{2007arXiv0711.1019G} Gaskell, C.~M., 
Goosmann, R.~W., Merkulova, N.~I., Shakhovskoy, N.~M., 
\& Shoji, M.\ 2007, arXiv:0711.1019

\bibitem[H{\"o}nig et 
al.(2006)]{2006A&A...452..459H} H{\"o}nig, S.~F., Beckert, T., Ohnaka, K., \& Weigelt, G.\ 2006, \aap, 452, 459 
\bibitem[H{\"o}nig \& Kishimoto(2009)]{2009arXiv0909.4539H} H{\" o}nig, S.~F., \& Kishimoto, M.\ 2009, arXiv:0909.4539

\bibitem[Jaffe et al.(2004)]{2004Natur.429...47J} Jaffe, W., et al.\ 2004, 
\nat, 429, 47 

\bibitem[Kishimoto et al.(2005)]{2005MNRAS.364..640K} Kishimoto, M., Antonucci, R., \& Blaes, O.\ 2005, \mnras, 364, 640 
\bibitem[Kishimoto et al.(2007)]{2007A&A...476..713K} Kishimoto, M., H{\"o}nig, S.~F., Beckert, T., \& Weigelt, G.\ 2007, \aap, 476, 713 
\bibitem[Kishimoto et al.(2008)]{2008Natur.454..492K} Kishimoto, M., Antonucci, R., Blaes, O., Lawrence, A., Boisson, C., Albrecht, M., \& Leipski, C.\ 2008, \nat, 454, 492
\bibitem[Kishimoto et al.(2009a)]{2009A&A...493L..57K} Kishimoto, M., H{\"o}nig, S.~F., Tristram, K.~R.~W., \& Weigelt, G.\ 2009, \aap, 493, L57 
\bibitem[Kishimoto et al.(2009b)]{2009A&A...507L..57K} Kishimoto, M., H{\"o}nig, S.~F., Antonucci, R., Kotani, T., Barvainis, R., Tristram, K.~R.~W., \& Weigelt, G.\ 2009, \aap, 507, L57 

\bibitem[Kobayashi et al.(1993)]{1993ApJ...404...94K} Kobayashi, Y., Sato, 
S., Yamashita, T., Shiba, H., \& Takami, H.\ 1993, \apj, 404, 94 
\bibitem[Koshida et al.(2009)]{2009ApJ...700L.109K} Koshida, S., et al.\ 
2009, \apjl, 700, L109

\bibitem[Lacy et al.(1982)]{1982ApJ...256...75L} Lacy, J.~H., et al.\ 1982, 
\apj, 256, 75 

\bibitem[Minezaki et al.(2004)]{2004ApJ...600L..35M} Minezaki, T., Yoshii, 
Y., Kobayashi, Y., Enya, K., Suganuma, M., Tomita, H., Aoki, T., 
\& Peterson, B.~A.\ 2004, \apjl, 600, L35
\bibitem[Mor et al.(2009)]{2009ApJ...705..298M} Mor, R., Netzer, H., 
\& Elitzur, M.\ 2009, \apj, 705, 298 
\bibitem[Mundell et al.(2003)]{2003ApJ...583..192M} Mundell, C.~G., Wrobel, 
J.~M., Pedlar, A., \& Gallimore, J.~F.\ 2003, \apj, 583, 192

\bibitem[Nagar\& Wilson(1999)]{1999ApJ...516...97N} Nagar, N.~M., \& Wilson, A.~S.\ 1999, \apj, 516, 97 
\bibitem[Nenkova et al.(2008a)]{2008ApJ...685..147N} Nenkova, M., Sirocky, 
M.~M., Ivezi{\'c}, {\v Z}., \& Elitzur, M.\ 2008a, \apj, 685, 147 
\bibitem[Nenkova et al.(2008b)]{2008ApJ...685..160N} Nenkova, M., Sirocky, 
M.~M., Nikutta, R., Ivezi{\'c}, {\v Z}., 
\& Elitzur, M.\ 2008b, \apj, 685, 160 
\bibitem[Netzer(2009)]{2009ApJ...695..793N} Netzer, H.\ 2009, \apj, 695, 
793 

\bibitem[Onken et al.(2007)]{2007ApJ...670..105O} Onken, C.~A., et al.\ 
2007, \apj, 670, 105 

\bibitem[Pacholczyk 
\& Weymann(1968)]{1968AJ.....73..870P} Pacholczyk, A.~G., \& Weymann, R.~J.\ 1968, \aj, 73, 870 

\bibitem[Raban et al.(2009)]{2009MNRAS.394.1325R} Raban, D., Jaffe, W., 
R{\" o}ttgering, H., Meisenheimer, K., \& Tristram, K.~R.~W.\ 2009, \mnras, 394, 1325 
\bibitem[Ragland et al.(2008)]{2008SPIE.7013E..10R} Ragland, S., et al.\ 
2008, \procspie, 7013,  
\bibitem[Rees et al.(1969)]{1969Natur.223..788R} Rees, M.~J., Silk, J.~I., 
Werner, M.~W., \& Wickramasinghe, N.~C.\ 1969, \nat, 223, 788
\bibitem[Riffel et 
al.(2006)]{2006A&A...457...61R} Riffel, R., Rodr{\'{\i}}guez-Ardila, A., \& Pastoriza, M.~G.\ 2006, \aap, 457, 61 
\bibitem[Riffel et al.(2009)]{2009ApJ...698.1767R} Riffel, R.~A., 
Storchi-Bergmann, T., \& McGregor, P.~J.\ 2009, \apj, 698, 1767 
\bibitem[Ruiz et al.(2003)]{2003MNRAS.340..733R} Ruiz, M., Young, S., 
Packham, C., Alexander, D.~M., \& Hough, J.~H.\ 2003, \mnras, 340, 733

\bibitem[Sanders et al.(1989)]{1989ApJ...347...29S} Sanders, D.~B., 
Phinney, E.~S., Neugebauer, G., Soifer, B.~T., 
\& Matthews, K.\ 1989, \apj, 347, 29
\bibitem[Schartmann et 
al.(2005)]{2005A&A...437..861S} Schartmann, M., Meisenheimer, K., Camenzind, M., Wolf, S., \& Henning, T.\ 2005, \aap, 437, 861
\bibitem[Schmitt 
\& Kinney(1996)]{1996ApJ...463..498S} Schmitt, H.~R., \& Kinney, A.~L.\ 1996, \apj, 463, 498 
\bibitem[Shapovalova et 
al.(2008)]{2008A&A...486...99S} Shapovalova, A.~I., et al.\ 2008, \aap, 486, 99
\bibitem[Smith et al.(2004)]{2004MNRAS.350..140S} Smith, J.~E., Robinson, 
A., Alexander, D.~M., Young, S., Axon, D.~J., 
\& Corbett, E.~A.\ 2004, \mnras, 350, 140
\bibitem[Suganuma et al.(2006)]{2006ApJ...639...46S} Suganuma, M., et al.\ 
2006, \apj, 639, 46
\bibitem[Swain et al.(2003)]{2003ApJ...596L.163S} Swain, M., et al.\ 2003, 
\apjl, 596, L163

\bibitem[Tristram et 
al.(2009)]{2009A&A...502...67T} Tristram, K.~R.~W., et al.\ 2009, \aap, 502, 67 

\bibitem[Wandel et al.(1999)]{1999ApJ...526..579W} Wandel, A., Peterson, 
B.~M., \& Malkan, M.~A.\ 1999, \apj, 526, 579
\bibitem[Wittkowski et 
al.(2004)]{2004A&A...418L..39W} Wittkowski, M., Kervella, P., Arsenault, R., Paresce, F., Beckert, T., \& Weigelt, G.\ 2004, \aap, 418, L39 
\bibitem[Wizinowich et al.(2006)]{2006SPIE.6268E..21W} Wizinowich, P., et 
al.\ 2006, \procspie, 6268,  














\end{thebibliography}
\end{document}